# Discoverability matters: Open access models and the translation of science into patents

Abdelghani Maddi[1*], Chongjun Xi[1], Xiaoting Chen[2], Isabelle Dorsch[1], Marc-André Simard[3,4]

[1] Sorbonne Université, CNRS, Groupe d'Étude des Méthodes de l'Analyse Sociologique de la Sorbonne, GEMASS, Paris, France

[2] School of Information Resources Management, Renmin University of China, Beijing 100872, China

[3] École de bibliothéconomie et des sciences de l'information, Université de Montréal, Montréal, Canada

[4] Centre interuniversitaire de recherche sur la science et la technologie (CIRST), Université du Québec à Montréal, Montréal, Canada

***corresponding author:** GEMASS, 59/61 rue Pouchet, 75017 Paris, France. Email: abdelghani.maddi@cnrs.fr

## Abstract

Scientific research is a key input into technological innovation, yet not all scientific knowledge is equally mobilized in patents. This paper examines how different scientific publishing models shape both the selection of scientific publications cited in patents and their cognitive alignment with patented technologies.

Using large-scale data on non-patent references linking patents to scientific publications, combined with metadata from OpenAlex, we compare the Open Access (OA) structure of patent-cited science to that of the scientific literature. We then assess cognitive alignment using semantic similarity between patent abstracts and the abstracts of cited publications, distinguishing between citations appearing in the front section of patents and those embedded in the body of patent texts.

We find that patent citations disproportionately draw on publications disseminated through highly visible and institutionally established publishing channels, particularly hybrid and bronze OA models, indicating strong selection effects. However, this dominance in citation counts does not translate into stronger cognitive alignment with patented technologies. On the contrary, publications in fully OA journals (gold and diamond OA) exhibit equal or higher semantic proximity, especially when cited in the body of patents.

These results suggest that the contribution of OA to innovation depends less on access alone than on how different publishing models are embedded in information infrastructures that shape the visibility, discoverability, and use of scientific knowledge.

## 1. Introduction

Scientific research is a central input into technological innovation [Coccia, 2020]. A large literature has documented how scientific knowledge contributes to invention, most notably through citations to scientific publications in patents, commonly referred to as non-patent references (NPRs) [Ahmadpoor & Jones, 2017; Marx & Fuegi, 2022; Maddi, 2024; Poege et al., 2019]. These citations have been widely used to study science–technology linkages, patterns of knowledge diffusion, and the economic impact of public research [Quemeneret et al., 2024]. At the same time, this literature has emphasized that patent citations are not neutral traces of knowledge flows, but are shaped by institutional routines, disclosure requirements, and information environments in which patents are produced [Verluise et al., 2025].

In parallel, scientific publishing has undergone profound transformations with the expansion of open access (OA). Promoted as a means to accelerate the circulation and use of scientific knowledge, OA has become a central component of science and innovation policy worldwide [Hopf et al., 2024]. However, OA is not a single model: it encompasses a variety of publishing arrangements (hybrid, bronze, green, gold, and diamond) that differ in governance, business models, editorial practices, and integration into indexing and discovery infrastructures [Piwowar et al 2018].

These OA models also differ in their practical implementation. Green OA refers to self-archiving in repositories; gold OA to immediate open access publication, typically involving article processing charges (APCs); hybrid OA to open access options within subscription journals; and bronze OA to free-to-read content without a clearly identifiable open license. Diamond OA generally refers to journals that provide immediate open access without charging APCs to authors. In this study, OA status is operationalized based on the definition provided by OpenAlex (see: https://help.openalex.org/hc/en-us/articles/24347035046295-Open-Access-OA).

Despite strong policy interest in open science as a driver of innovation, empirical evidence on the relationship between OA and technological development remains mixed [Hopf et al., 2024]. Existing studies have primarily examined whether open access and its models increase citations, visibility, or diffusion within science, often treating them as a binary attribute, although some work has explored these distinctions in greater depth. Much less is known about how different OA models shape the use of scientific knowledge beyond academia, particularly in patents [Jahn et al., 2022].

Multiple studies confirm that openness (through open innovation strategies, open data policies, or trade liberalization) can lower barriers to technology adoption by increasing access to external knowledge and resources (Cui et al., 2022; Xu, 2024; Schlagwein et al., 2017; Gonçalves et al., 2021; Zuiderwijk et al., 2015; Cao et al., 2024). For example, open innovation approaches enable organizations to source knowledge externally and leverage IT tools for both internal and external information flows (Cui et al., 2022). Trade openness has been shown to facilitate technology transfer and productivity growth in high- and middle-income countries (Gonçalves et al., 2021; Iftikhar et al., 2025). This access-based perspective overlooks two important issues. First, not all OA models offer the same degree of visibility and integration into the information systems used by inventors, patent examiners, and patent offices. Second, citation in a patent does not necessarily imply substantive use of scientific knowledge; some references reflect formal disclosure rather than cognitive integration into the inventive process (Shu, al., 2026).

These limitations point to a more fundamental question: do publishing models primarily affect which scientific publications are cited in patents, or do they also shape how closely cited science aligns with patented technologies? In other words, publishing models may influence the selection of scientific publications into patent citations without necessarily affecting their translation into technological content.

Highly visible and institutionally established publishing channels, often associated with traditional subscription journals and hybrid or bronze OA, may be overrepresented in patent citations because they are more easily discoverable within existing search tools (ex. international databases like Web of Science and Scopus), are embedded in professional routines, and are often perceived as more prestigious. Conversely, fully OA journals, including gold and diamond OA, may publish research that is equally or more relevant to technological invention, yet be less frequently cited if they are less embedded in dominant discovery infrastructures and perceived as less prestigious.

Distinguishing between selection and translation is essential for understanding the role of OA in innovation. Without this distinction, patterns of patent citation risk being misinterpreted as indicators of technological relevance rather than artifacts of visibility and institutional embedding.

Building on this perspective, this paper addresses two main research questions. First, how does the distribution of OA models among scientific publications cited in patents compare to that of the scientific literature as a whole, once disciplinary differences are taken into account? Second, conditional on being cited, do publications disseminated through different OA models differ in their cognitive alignment with patented technologies?

To answer these questions, we further distinguish between citations appearing in the front section of patents and those embedded in the body of patent texts, as these positions are commonly associated with different functions and degrees of substantive knowledge use. We hypothesize that traditional and highly visible publishing channels will be overrepresented among patent citations, reflecting selection effects, but that this overrepresentation will not necessarily correspond to higher cognitive alignment with patented technologies, particularly for citations embedded in the body of patents.

This study empirically tests a set of hypotheses derived from a conceptual framework that views the incorporation of scientific knowledge into technological innovation as a two-stage process involving selection and translation. First, we examine whether the distribution of OA publishing models among scientific publications cited in patents deviates from that observed in the broader scientific literature, reflecting systematic biases in knowledge discovery and retrieval. Second, we investigate whether the OA models that are most frequently cited are also those that exhibit the strongest cognitive alignment with patent content, thereby assessing whether visibility translates into substantive technological relevance. Finally, we analyze whether the relationship between OA models and cognitive alignment varies according to the functional location of citations within patents, distinguishing between inventor-driven body citations and administratively introduced front-section citations. Together, these hypotheses allow us to move from descriptive citation patterns to the underlying mechanisms shaping the translation of scientific knowledge into technological applications.

This paper makes three contributions. First, it brings the institutional diversity of OA models into the analysis of science–technology linkages, moving beyond binary distinctions between open and closed access. Second, it introduces an explicit distinction between selection into patent citations and cognitive alignment with invention, helping reconcile mixed findings on OA and innovation. Third, it highlights the role of discoverability (the embedding of publishing models in information infrastructures) as a key mechanism linking scholarly communication to technological development.

## 2. Background and related literature

### 2.1. Measuring Science-Technology Linkage

Understanding how scientific knowledge contributes to technological innovation has long been a central concern in the economics of innovation and scientometrics. Patent citations to scientific publications, commonly referred to as non-patent references (NPRs), have become the dominant empirical tool for tracing science–technology linkages since the seminal contributions of Carpenter et al. (1983) and Narin et al. (1985). Subsequent work has shown that patents, particularly in science-intensive fields, draw heavily on publicly funded research, highlighting the foundational role of science in technological development (Narin et al., 1997; Ahmadpoor et al., 2017; Fleming et al., 2019).

At the same time, substantial literature has emphasized the limits of interpreting patent citations as direct measures of knowledge flows. Citations may serve multiple functions, ranging from substantive acknowledgment of prior scientific input to legal disclosure or background contextualization (Meyer, 2000). Moreover, the distinction between citations introduced by patent applicants and those added by examiners has generated sustained debate. While some studies argue that examiner citations introduce noise unrelated to the inventor's cognitive process (Jaffe et al., 2000), others suggest that examiner-added references often reflect highly relevant prior art and technical proximity (Criscuolo et al., 2008; Chen et al., 2017).

These debates point to a critical limitation of citation-based approaches: citation counts alone cannot distinguish between the selection of scientific publications into patents and the depth of their substantive integration into technological inventions. As a result, understanding science–technology linkages requires moving beyond the mere presence of citations toward indicators that capture how scientific knowledge is used.

## 2.2. Open access, visibility, and the selection of science into patents

In parallel to these developments, the ecosystem of scholarly communication has been transformed by the rise of OA. Open access has been widely promoted as a means to accelerate the diffusion of scientific knowledge and enhance its societal and economic impact by lowering access barriers. From this perspective, easier access to scientific publications should facilitate their use by firms and inventors, particularly those with limited resources.

Empirical evidence lends partial support to this view (Maddi et al. 2024). Several studies document a rising share of OA publications among patent citations, especially in science-based sectors such as biomedicine (Jahn et al., 2022). Other work shows that OA publications are overrepresented in patent citations relative to their share in the scientific corpus, suggesting that accessibility plays a role in industrial knowledge use (Maddi et al., 2024). More fine-grained analyses indicate that some OA models, such as hybrid or green OA, are more likely to be cited by patents than others (Dorta-González et al., 2025), and that informal access channels further amplify this advantage.

However, this literature largely interprets these patterns through an access-based lens, implicitly equating higher citation probability with greater technological relevance. This interpretation overlooks two important considerations. First, OA is not a homogeneous condition: hybrid, bronze, green, gold, and diamond OA differ substantially in their integration into established publishing platforms, indexing systems, and discovery tools, with important implications for equity in access, visibility, and participation across regions and institutions (Simard et al., 2025; Maddi et al., 2024; Piwowar et al., 2018; Simard et al., 2024). Second, patent citation practices are embedded in professional routines and infrastructures that privilege visibility and discoverability, not only legal access. Highly established journals and publishers, often associated with hybrid or bronze OA, may be more easily surfaced in search processes used by inventors and patent offices, independently of the substantive relevance of their content.

From this perspective, OA models may shape the selection of scientific publications into patents primarily through differences in discoverability rather than through differences in the underlying technological relevance of the research they disseminate.

### 2.3. From selection to translation: measuring cognitive alignment through text similarity

While existing studies convincingly show that OA affects whether scientific publications are cited in patents, much less is known about whether it affects how scientific knowledge is translated into technological content. Citation-based measures cannot address this question directly, as they do not capture the degree to which inventors engage with, understand, or reuse the content of cited publications.

To address this limitation, recent research has increasingly turned to Natural Language Processing (NLP) and text-based methods. Early work by Chen et al. (2017) demonstrated that patents and the scientific publications they cite exhibit higher textual similarity than non-citing pairs, validating text similarity as a proxy for cognitive proximity. Subsequent studies have extended this approach to capture different dimensions of science–technology interaction, including novelty, recombination, and the role of scientific language in invention (Comai et al., 2018; Denter et al., 2025).

Text-based similarity measures offer a crucial advantage: they allow researchers to distinguish between being cited and being substantively aligned with patented technologies. High semantic similarity suggests that scientific concepts, terminology, or problem framings have been incorporated into the inventive process, whereas low similarity may indicate more peripheral or formal citation.

Despite these advances, no existing study has systematically combined OA models with semantic similarity analysis of patent–science linkages. As a result, a key question remains unanswered: does OA, and the diversity of its publication models, affect not only the likelihood that scientific publications are cited in patents, but also the depth of their cognitive integration into technological invention?

### 2.4. Research gap and contribution

Bringing these strands together, the literature reveals a clear gap. We know that science matters for innovation, that patent citations imperfectly trace this relationship, and that OA increases the visibility of scientific publications in patents. We also know that text-based methods can capture cognitive proximity between science and technology. What remains missing is an integrated analysis that examines how different OA models shape both the selection of scientific publications into patents and their translation into technological content. In other words, this study will move beyond previous research that only explored whether scientific papers are "more likely to be cited in patents" and go further to empirically testing whether these cited papers are "truly absorbed by technology".

This paper addresses this gap by explicitly distinguishing between selection and cognitive alignment, and by analyzing how these two dimensions vary across OA models and citation positions within patents. In doing so, it shifts the focus from access alone to the role of discoverability and information infrastructures in shaping the pathways through which scientific knowledge contributes to technological innovation.

### 2.5. Hypotheses

This paper conceptualizes the flow of scientific knowledge into the technological domain as a two-stage process consisting of selection and translation. In the selection stage, scientific publications must first be discovered and identified by inventors or patent examiners within the vast information landscape, a process heavily mediated by visibility and search costs. In the translation stage, the selected knowledge is cognitively integrated into the technological content of the patent, requiring a degree of semantic alignment between the two domains. We argue that the distribution of OA models within patent citations deviates from the broader scientific landscape because patent citation practices are not random but are deeply embedded in specific search routines and information infrastructures. While the general scientific literature reflects a diverse array of publishing models, patent systems rely on professional databases (such as EPOQUENet or Lens.org) and institutionalized search protocols that prioritize reliability and ease of access. Within this framework, hybrid OA publications benefit from a "prestige-accessibility" dual advantage: they are hosted in established, high-reputation subscription journals that signal authority to examiners while bypassing paywall barriers. Similarly, bronze OA publications, despite their lack of open licenses, are typically hosted on large-scale commercial publisher platforms that are seamlessly integrated into patent office search infrastructures through standardized metadata and persistent identifiers. Consequently, these models achieve a level of institutional visibility that newer gold OA journals or fragmented green OA repositories often lack. Based on this systematic bias in the discovery and retrieval process, we propose the following:

**H1.** *The distribution of OA models among scientific publications cited in patents differs from that of the scientific literature, with hybrid and bronze OA publications being overrepresented.*

Building upon the selection bias identified in the first stage, it is crucial to examine whether this heightened visibility translates into a substantive contribution to technological innovation. This paper posits a potential decoupling between the frequency of selection and the depth of knowledge translation. Although hybrid and bronze OA models enjoy an "overrepresentation" due to their institutional visibility and integration into search infrastructures, this advantage is often driven by the standardized search routines of patent examiners rather than the intrinsic research needs of inventors. Such citations may reflect administrative compliance rather than the genuine cognitive absorption of scientific knowledge. In contrast, scientific works that are more closely aligned with practical technological problem-solving, often found in green OA repositories, may face higher search frictions but exhibit stronger technical relevance. Therefore, cognitive alignment, which measures the semantic proximity between scientific and technological content, is not necessarily proportional to citation frequency driven by discoverability. If certain publishing models are overrepresented primarily due to their "ease of discovery," they may not demonstrate superior cognitive alignment, and may even show lower average similarity due to the inclusion of more peripheral or perfunctory citations. Based on this logic, we propose:

**H2.** *Open access models that are more frequently cited in patents are not necessarily associated with higher cognitive alignment between scientific publications and patents.*

Finally, the positioning of citations within a patent document reflects distinct functions of scientific references and acts as a crucial moderator in the relationship between OA models and cognitive alignment. Citations embedded in the body of the patent are predominantly driven by the inventors themselves during the R&D process, representing a substantive engagement with scientific knowledge to solve technical problems. Conversely, citations in the front-page section are more frequently introduced by patent attorneys or examiners during prior-art searches to fulfill legal disclosure requirements and define the scope of claims (Marx & Fuegi, 2020). Thus, we argue that because body citations stem from genuine cognitive interaction, their semantic content is more likely to resonate with the patent's technological disclosures. Consequently, while different OA models shape the visibility of publications through information infrastructures, the impact of these models on cognitive alignment should be more pronounced and sensitive in inventor-driven body citations. In contrast, the administrative and procedural nature of front-section citations may introduce "noise" that flattens the inherent differences in knowledge translation efficiency across various OA types. Based on this functional distinction, we propose:

**H3.** *Differences in cognitive alignment across OA models are stronger for body citations than for front-section citations.*

In summary, these three hypotheses form a cohesive analytical chain that moves from surface-level patterns to underlying mechanisms. H1 (The Visibility Hypothesis) establishes the empirical phenomenon of citation bias, attributing the dominance of specific OA models to their integration into established information infrastructures. H2 (The Alignment Hypothesis) challenges the assumption that citation frequency equals knowledge impact, decoupling "being seen" from "being substantively used." Finally, H3 (The Structural Validation) serves as a critical test of this decoupling; using the functional difference between front-page and body citations, it provides a more granular lens to verify whether the observed citation patterns are driven by professional search routines or genuine technological translation.

### 3. Data and Methods

#### 3.1. Data sources: patents, non-patent references, and scientific publications

Our analysis builds on the comprehensive dataset of non-patent references (NPRs) developed by Marx and Fuegi (2020), which links patents to scientific publications cited as prior art. The full dataset contains 44,778,211 NPRs and provides detailed information on the position of each reference within the patent document (front section, body, or both), the type of citing actor (applicant or examiner), and a confidence score assessing the quality of the publication–patent match. Each NPR is also linked to a corresponding OpenAlex identifier, allowing systematic retrieval of publication metadata.

Across the full dataset, 27.8% of NPRs appear in the body of patents (12,459,253 references), 64.2% in the front section (28,727,216 references), and 8.0% in both locations (3,591,742 references). This positional information is central to our analysis, as citation position is commonly associated with different functions and actors in the patenting process.

Metadata on scientific publications (including abstracts, disciplinary classification, OA status, publication year, and normalized citation impact) were retrieved from OpenAlex, by using the R (V4.4.1) package openalexR (V2.0.2).

Patent abstracts, enabling text-based comparison between patents and cited publications, were obtained from the PatentsView (USPTO) database (https://patentsview.org/) merged with Marx and Fuegi's data by using the patents identifiers.

### 3.2. Construction of the analytical samples

The datasets corresponding to publications cited exclusively in the front section or exclusively in the body of patents are substantially larger, containing more than 28 million and 12 million NPRs respectively. For computational and analytical tractability, we adopt a sampling strategy designed to preserve representativeness while enabling direct comparison across citation positions.

To assess the robustness of results to sampling, we first draw ten independent random samples of 50,000 publications from the 2,015,841 publication–patent pairs in the core dataset and compare their distributions in terms of OA status and semantic similarity scores (see Appendix A1 for descriptive statistics). Results reported in the Appendix A2 show that variation across samples is minimal and statistically insignificant.

Based on this validation, we apply the same sampling approach to the front and body NPR datasets. For each citation position (front, body, both), we construct a random sample of 50,000 distinct scientific publications meeting the following criteria: (i) the publication has an abstract, (ii) the citing patent has an abstract, and (iii) the publication can be matched to OpenAlex metadata.

The final analytical sample therefore consists of three datasets of equal size (50,000 publications each), corresponding to front, body, and both citation positions. For each dataset, we compute publication–patent semantic similarity scores and retrieve OA status, disciplinary classification, publication year, and normalized citation impact.

### 3.3. Baseline distribution of scientific publications

To establish a reference point for the analysis, we construct a baseline distribution of scientific publications using OpenAlex. Specifically, we extract the counts of all research articles indexed in OpenAlex and compute their distribution by OA status and broad disciplinary domain. This baseline represents the OA structure of published science against which the composition of patent-cited publications can be meaningfully compared.

Because publishing practices and OA adoption vary substantially across scientific domains, all comparisons between patent-cited publications and the baseline are normalized by domain. This domain-level normalization ensures that observed differences in OA composition are not driven by field-specific publication patterns, but instead reflect selection mechanisms associated with patent citation practices.

**Table 1. OpenAlex baseline distribution of research articles by OA status and domain** *(Shares within domain (percent of research articles, with absolute counts))*

| Domain | Closed | Gold OA | Green OA | Bronze OA | Diamond OA | Hybrid OA | Total (N) |
|---|---|---|---|---|---|---|---|
| Life Sciences | 65.8% (16,282,537) | 8.0% (1,974,937) | 7.9% (1,968,017) | 7.6% (1,891,880) | 6.5% (1,616,986) | 4.2% (1,027,669) | 24,762,026 |
| Health Sciences | 69.3% (27,978,443) | 6.9% (2,805,038) | 6.0% (2,410,310) | 7.2% (2,912,434) | 7.2% (2,901,554) | 3.4% (1,387,879) | 40,395,658 |
| Physical Sciences | 74.5% (51,378,948) | 5.0% (3,419,232) | 7.4% (5,128,013) | 4.7% (3,207,583) | 5.7% (3,962,828) | 2.7% (1,839,192) | 68,935,796 |
| Social Sciences | 71.1% (42,807,749) | 3.0% (1,832,202) | 10.2% (6,140,321) | 3.2% (1,913,209) | 9.5% (5,687,432) | 3.0% (1,798,232) | 60,179,145 |

Table 1 reports the baseline distribution of research articles by OA status across major scientific domains. Across all domains, closed-access publications remain dominant, accounting for between 66% and 75% of research articles. However, the composition of OA models varies substantially across domains. Life and health sciences exhibit relatively higher shares of gold and bronze OA, whereas physical sciences remain more heavily oriented toward closed access. Social sciences display a distinct profile, with comparatively higher shares of green and diamond OA. This heterogeneity underscores the importance of controlling for disciplinary composition when comparing the OA structure of patent-cited publications to that of the scientific literature.

This baseline distribution serves two analytical purposes. First, it provides the reference against which the OA composition of patent-cited science is evaluated in the analysis of selection effects (H1). Second, it allows subsequent analyses of cognitive alignment to be interpreted conditional on the underlying OA landscape of published research within each domain.

### 3.4. Text Similarity Measurement

To assess cognitive alignment between scientific publications and patents, we compute semantic similarity scores based on textual representations of abstracts. For each publication–patent pair, we embed the abstract of the scientific publication and the abstract of the citing patent into a shared vector space using a pre-trained language model. To quantify the linkage strength between patent technology and scientific knowledge, this study adopts the classic Vector Space Model (VSM) framework (Salton et al., 1975). Magerman et al. (2010) showed that VSM-based text mining techniques effectively detect textual similarity between patent documents and scientific publications. Furthermore, Ahlgren et al. (2009) noted that VSM-based techniques possess higher accuracy compared to other document similarity metrics. The specific calculation process of this study involves three steps: model selection, text vectorization, and similarity calculation.

(1) Model Selection: Traditional bag-of-words models or general pre-trained models (e.g., BERT) often struggle to precisely capture deep semantic associations within specific disciplines. Therefore, we selected the SPECTER 2 model developed by the Allen Institute for AI (see https://github.com/allenai/SPECTER2). Based on a "citation-informed" training objective, this model is capable of mapping documents with citation relationships into proximal vector spaces. Although SPECTER 2 is primarily trained on scientific literature corpora, it is highly feasible for patent text embedding. First, patents are essentially carriers of technical solutions; in fields with high scientific relevance such as artificial intelligence and biomedicine, patent texts share a vast amount of core technical terminology with academic papers, objectively constructing an overlapping semantic space. Second, given that this study aims to measure the "scientific nature" of patents and their distance from frontier science, projecting patent texts into a feature space defined by massive scientific publications intuitively reflects their positioning in the scientific dimension. Furthermore, from a technical implementation perspective, the Transformer-based attention mechanism endows the model with excellent robustness. It effectively reduces the attention weight assigned to generic patent legal boilerplate, focusing instead on substantive technical semantics, thereby ensuring the accuracy of cross-domain representation learning.

(2) Text Vector Embedding: For each established "patent-paper" pair, we extracted the abstract text of the patent and the abstract text of the paper, respectively. Using the SPECTER 2 model, these unstructured texts were transformed into fixed-dimension dense vectors. This process achieves semantic alignment across text genres, represented as:

$$V_{pat} = \text{Encoder}(T_{pat}), V_{pub} = \text{Encoder}(T_{pub})$$

Where $V_{pat}$ and $V_{pub}$ denote the embedding vectors (dimension d = 768) for the patent and the paper, respectively.

(3) Similarity Calculation: Upon obtaining the vector representations, we employed Cosine Similarity to measure the proximity of their semantic content. Compared to Euclidean distance, Cosine Similarity focuses on directional consistency within the semantic space, effectively mitigating differences in abstract length and writing style between patents and papers. The calculation formula is as follows:

$$Sim(p,l) = \frac{V_{pat} \cdot V_{pub}}{\|V_{pat}\| \|V_{pub}\|}$$

The value ranges from [-1, 1]. A value closer to 1 indicates that the patent and the cited paper are more similar in research topics and technical content, implying a closer "knowledge distance."

### 3.5. Analytical strategy

The analysis proceeds in two main steps. First, we examine selection effects by comparing the OA composition of scientific publications cited in patents to the OA distribution observed in the scientific corpus, controlling for disciplinary domains. This analysis is conducted separately for front, body, and both citation positions.

Second, we analyze cognitive alignment by examining differences in semantic similarity scores across OA models. We estimate descriptive statistics, distributional comparisons, and multivariate regression models that relate semantic similarity to OA status, citation position, disciplinary domain, publication year, and normalized citation impact (FWCI).

Additional models including interaction terms between OA models and disciplinary domains, as well as between OA models and citation positions, are reported in the Appendix A4. These analyses assess the contextual dependence of observed associations and serve as robustness checks.

### 4. Result

### 4.1. Selection effects: open access models in patent-cited science (H1)

We begin by examining whether the OA composition of scientific publications cited in patents differs from that of the scientific literature as a whole. This analysis directly tests Hypothesis H1, which posits that patent citations reflect selection processes shaped by publishing models and information infrastructures rather than a representative sampling of published science.

Figure 1 compares the OA composition of scientific publications cited in patents to that of the scientific literature as a whole, using OpenAlex as a baseline. For each disciplinary domain and citation position (Front, Body, Both), the figure reports log2 enrichment scores, computed as the ratio between the share of a given OA model in the patent-cited dataset and its corresponding share in the OpenAlex baseline, within domain (enrichment is expressed as log2(observed share ÷ baseline share). Values of +1 and −1 correspond respectively to a doubling or a halving relative to the baseline). Positive values therefore indicate over-representation in patent citations relative to expectations based on the structure of scientific publishing, while negative values indicate under-representation.

**Figure 1. OA status enrichment in patent-cited publications compared to the OpenAlex baseline (within domains).**

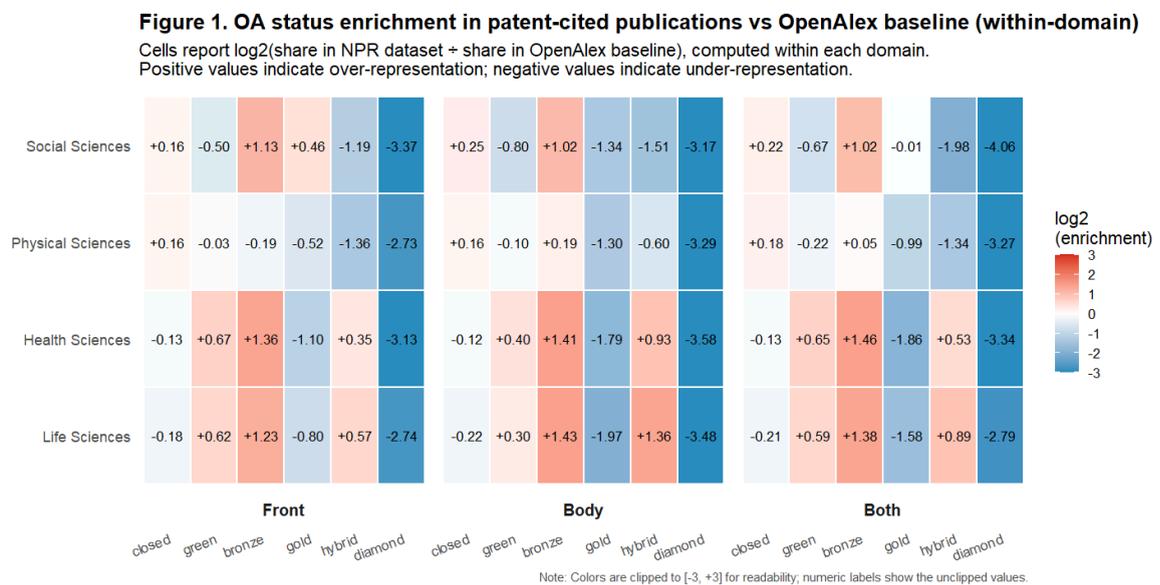

A first striking result is the systematic over-representation of bronze and hybrid OA across most domains and citation positions. This pattern is particularly pronounced in the Life and Health Sciences, where bronze OA exhibits strong positive enrichment in both front-section and body citations. Hybrid OA also shows consistent positive enrichment, especially for body and combined citations. By contrast, diamond OA is sharply under-represented in all domains and citation positions, with large negative enrichment values that are remarkably stable across fields. This under-representation persists even in Life Sciences, where diamond OA constitutes a non-negligible share of the baseline scientific corpus.

Gold OA displays a more nuanced pattern. Across domains, it tends to be under-represented or only weakly represented in patent citations, especially in Physical and Social Sciences. While some positive enrichment appears in specific domain–position combinations, gold OA never reaches the levels observed for hybrid or bronze OA. Closed access publications, by contrast, are generally close to parity with the baseline, suggesting neither strong over- nor under-selection once domain composition is taken into account.

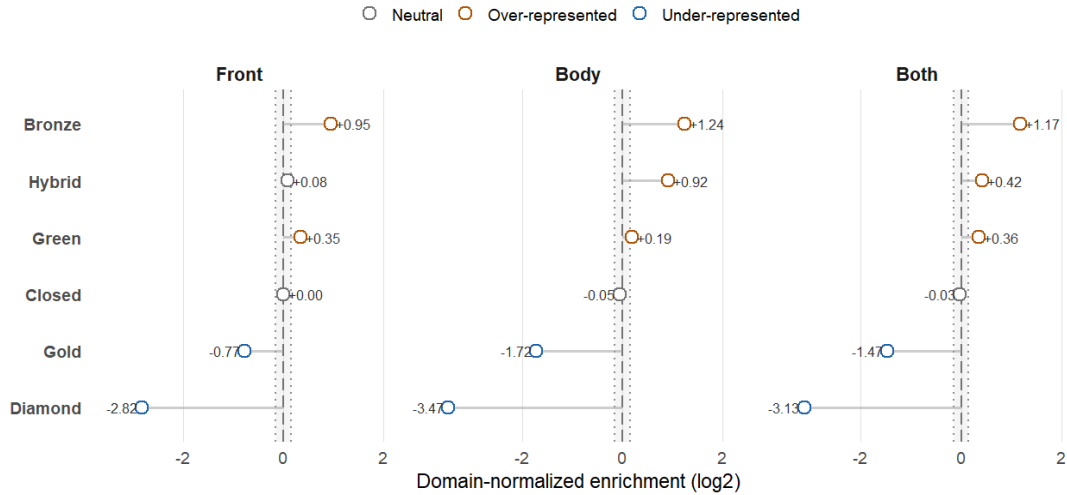

Figure 2. OA models are not equally represented in patent-linked publications
Domain-normalized enrichment by OA status (log2 scale). Shaded area indicates |enrichment| < 0.15.

Figure 2 synthesizes the within-domain selection patterns documented in Figure 1 by reporting a single, domain-normalized enrichment statistic for each OA model and citation position. Specifically, the figure plots the $\log_2$ ratio between the observed share of each OA status in the NPR datasets and the expected share given each dataset's domain mix. Values to the right of zero indicate over-representation among patent-linked publications relative to the baseline, while values to the left indicate under-representation; the shaded band around zero highlights small deviations treated as approximately neutral.

Consistent with H1, the results show strong and systematic selection effects that depend on citation position. Bronze OA is robustly over-represented across all positions (+0.95 in Front; +1.24 in Body; +1.17 in Both), while diamond OA is sharply under-represented in every dataset (−2.82 in Front; −3.47 in Body; −3.13 in Both). Hybrid OA exhibits a clear positional gradient: it is close to neutral in Front citations (+0.08) but becomes substantially over-represented for Body and Both citations (+0.92 and +0.42, respectively). In contrast, gold OA is consistently under-represented, particularly for Body and Both citations (−1.72 and −1.47). Green OA is modestly over-represented (+0.19 to +0.36), whereas closed access remains close to neutral by construction.

Overall, Figure 2 reinforces the key message of the selection analysis: patent-linked publications disproportionately come from high-visibility and institutionally embedded dissemination channels, but this pattern is not uniform across citation positions. The strongest asymmetries are observed for bronze and diamond OA, while hybrid OA becomes more prevalent precisely in citation contexts that are more likely to reflect substantive engagement with scientific knowledge (Body and Both).

### 4.2. Open access models and cognitive alignment with patented technologies (H2)

While Section 4.1 documents selection effects in patent citation practices, it does not address whether the publishing models most frequently cited in patents are also those most closely aligned with patented technologies. We now turn to this question by examining semantic similarity between patent texts and cited scientific publications. This analysis directly tests Hypothesis H2, which predicts a decoupling between citation frequency and cognitive alignment.

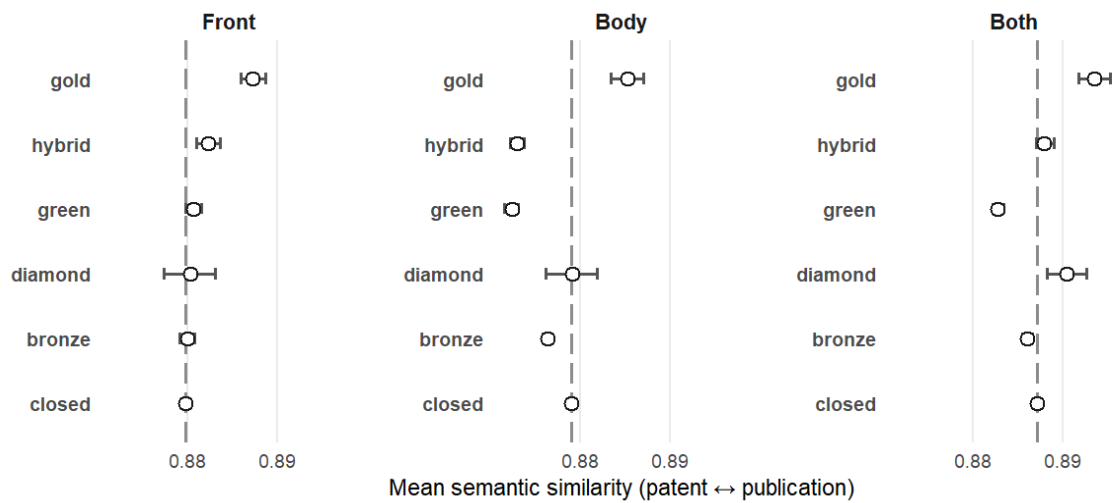

Figure 3 examines whether the over- and under-representation of OA models in patent citations documented in Figures 1 and 2 translates into differences in cognitive alignment between scientific publications and patented technologies. For each citation position (Front, Body, Both), the figure reports the mean semantic similarity between patent abstracts and the abstracts of cited publications, together with bootstrap 95% confidence intervals. To facilitate interpretation, the dashed vertical line in each panel indicates the mean similarity for closed-access publications within the same citation position.

Across citation positions, differences in average semantic similarity across OA models are moderate in magnitude but highly structured. In front-section citations, semantic similarity is relatively homogeneous across OA models, with most estimates clustering closely around the closed-access benchmark. Gold OA publications exhibit slightly higher average similarity than closed-access publications, while hybrid and green OA are marginally lower. Diamond OA displays wider confidence intervals, reflecting greater heterogeneity, but its mean similarity remains close to the closed-access reference.

Clearer differentiation emerges for body citations, which are more likely to reflect substantive engagement with scientific content. In this context, gold OA publications exhibit the highest average semantic similarity, exceeding that of closed-access publications by a non-trivial margin. Hybrid and green OA also perform comparably to, or slightly above, the closed-access benchmark, while bronze OA shows marginally lower similarity. Diamond OA remains close to the reference level but continues to display higher dispersion. These patterns indicate that publications disseminated through fully open models are, on average, at least as cognitively aligned with patented technologies as publications disseminated through traditional channels.

The pattern is reinforced for publications cited in both the front section and the body of patents. In this subset, which likely captures publications playing a central role in the inventive process, gold OA, and to a lesser extent diamond OA, exhibit the highest semantic proximity to patented technologies, clearly exceeding the closed-access benchmark. By contrast, hybrid and bronze OA cluster closer to the reference line. Consequently, these results provide direct support for H2 and H3: the OA models that are most overrepresented in patent citations are not those associated with the strongest cognitive alignment, and differences across OA models become more pronounced precisely in citation contexts that signal deeper knowledge integration.

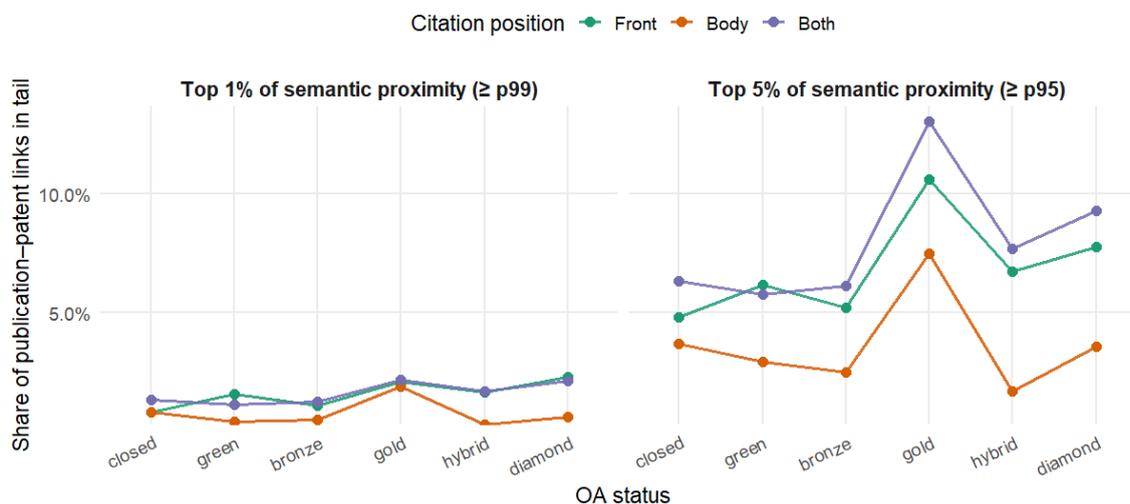

**Figure 4. Concentration of OA models in the highest semantic proximity links**
Tail shares computed using global thresholds: p95 = 0.925, p99 = 0.94

Figure 4 examines whether specific OA models are disproportionately represented among the most cognitively aligned patent–publication links. Rather than focusing on average semantic similarity, the figure reports the share of publication–patent links falling into the top 5% and top 1% of the global semantic similarity distribution, using thresholds computed across all datasets (p95 = 0.925; p99 = 0.94). This tail-based approach isolates cases of exceptionally strong cognitive proximity, which are more likely to reflect intensive reuse of scientific knowledge in the inventive process.

Two clear patterns emerge. First, gold OA publications consistently exhibit the highest concentration in the upper tail of semantic similarity, across citation positions and thresholds. This pattern is particularly pronounced for publications cited in both the front section and the body of patents, where gold OA accounts for the largest share of links in the top 5% and top 1% of similarity scores. Diamond OA shows a similar, though more heterogeneous, pattern: while its average representation is more variable, it systematically outperforms closed, hybrid, and bronze OA in the extreme upper tail, especially for front and combined citations.

Second, hybrid and bronze OA, despite their over-representation in patent citations overall, do not dominate the highest semantic proximity links. Their tail shares are consistently lower than those of gold OA and often comparable to, or below, those of closed-access publications. This contrast is particularly stark in the body-citation dataset, where hybrid and bronze OA display relatively low participation in the top 1% of semantic similarity, while gold OA remains clearly dominant.

Together, these results reinforce the conclusions drawn from Figure 3. Differences across OA models are modest when assessed in terms of mean semantic proximity, but become substantially more pronounced when attention is restricted to the most cognitively intensive patent–science links. The concentration of gold, and to a lesser extent diamond, OA in the extreme right tail provides strong support for H3, indicating that fully open publishing models are more likely to host scientific outputs that are deeply integrated into patented technologies. At the same time, the results further corroborate H2, showing that the prominence of hybrid and bronze OA in patent citation counts reflects selection and visibility effects rather than superior cognitive alignment.

### 4.3. Citation position and substantive use of scientific knowledge (H3)

We next examine whether differences in cognitive alignment across OA models depend on citation position within patents. Figure 5 examines where OA models differ from closed-access publications across the entire distribution of semantic similarity, using shift functions that report quantile-by-quantile differences relative to closed access. Positive values indicate that, at a given quantile, publications published under a given OA model exhibit higher semantic proximity to patents than closed-access publications cited in the same position.

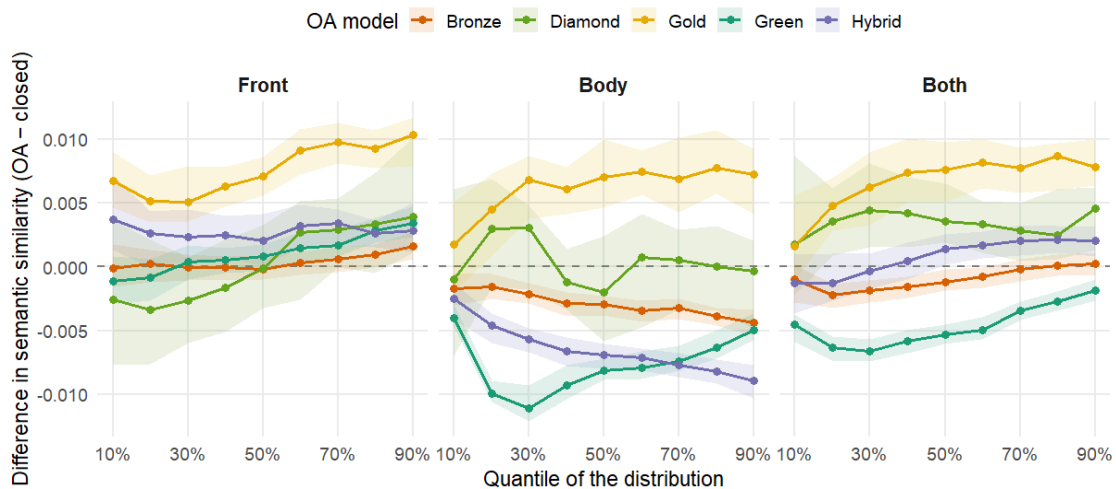

Figure 5. Where do OA models differ from closed? (distribution shift)
Shift functions: quantile differences in semantic similarity (OA − closed), with bootstrap 95% intervals.

A first important result is that differences across OA models are highly heterogeneous across citation positions, reinforcing the idea that citation position captures meaningful variation in the substantive use of scientific knowledge. For front-section citations, distributional shifts are generally modest and concentrated in the upper half of the distribution. Gold OA exhibits a clear and monotonic upward shift from the median onwards, while hybrid and green OA show smaller positive deviations. Bronze OA remains close to the closed-access benchmark throughout the distribution, and diamond OA displays wider uncertainty, suggesting greater heterogeneity rather than systematic advantage.

In contrast, body citations reveal pronounced and structured distributional shifts. Gold OA shows consistently positive differences across nearly the entire distribution, with the largest gains emerging from the median to the upper quantiles. This indicates that gold OA publications are not only more likely to appear among the most cognitively aligned patent–publication links, but that their entire similarity distribution is shifted upward relative to closed access. By contrast, green and hybrid OA display negative shifts across most quantiles, particularly in the lower and middle parts of the distribution, suggesting weaker cognitive alignment when citations reflect substantive engagement. Bronze OA remains slightly below the closed-access reference across quantiles, while diamond OA oscillates around zero with moderate positive deviations in the upper tail.

The pattern is most pronounced for publications cited in both the front section and the body of patents, which plausibly capture the most central scientific inputs to invention. In this subset, gold OA exhibits the strongest and most persistent upward shifts across all quantiles, with particularly large advantages in the upper half of the distribution. Diamond OA also displays positive shifts at higher quantiles, while hybrid and bronze OA cluster close to zero or remain slightly negative. These results align closely with the tail concentration patterns documented in Figure 4 and provide strong distributional evidence in support of H3: fully open publication models, especially gold OA, are disproportionately associated with the most substantively integrated science–technology linkages, whereas models that dominate patent citation counts do not exhibit comparable cognitive advantages.

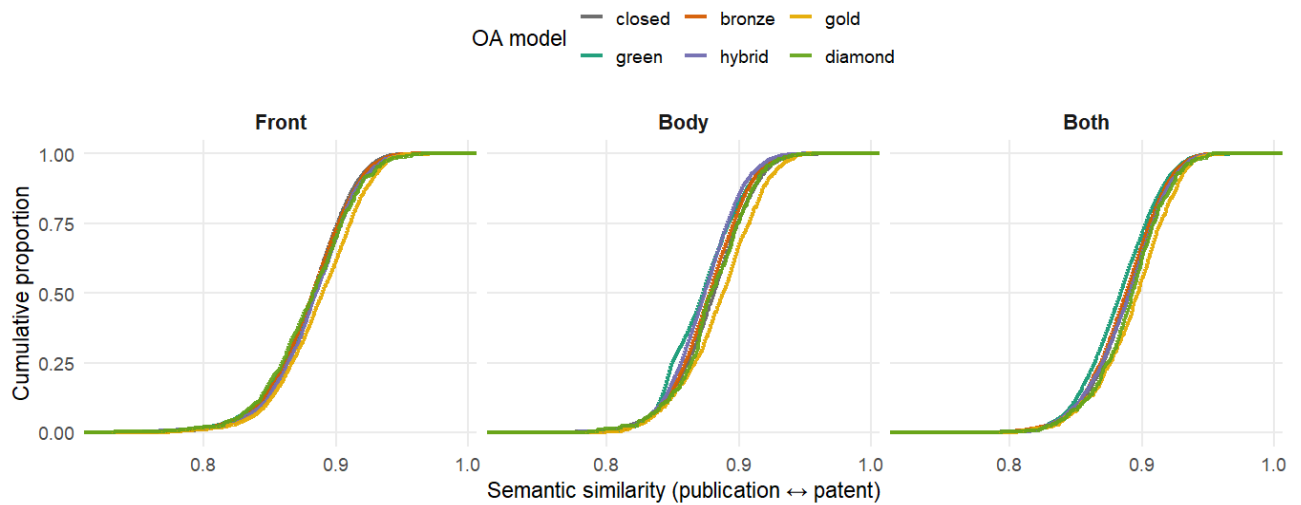

**Figure 6. Distribution of semantic similarity by OA model and citation position**
Empirical cumulative distribution functions (ECDFs) highlight systematic distribution shifts, especially in the upper tail

Figure 6 presents empirical cumulative distribution functions (ECDFs) of semantic similarity between patents and cited publications, disaggregated by OA model and citation position. ECDFs provide a non-parametric and distribution-wide perspective, allowing direct comparison of entire similarity distributions rather than relying on averages or tail thresholds. For a given citation position, a curve lying systematically to the right (or equivalently, lower at high similarity values) indicates first-order stochastic dominance and a greater concentration of high-similarity links.

Across all citation positions, gold OA exhibits a consistent rightward shift of the similarity distribution relative to closed, hybrid, and bronze OA, indicating a higher probability of observing strong semantic alignment with patented technologies. This dominance is most visible in the upper tail of the distribution, where the gold OA ECDF rises more slowly, reflecting a larger mass of publication–patent pairs with very high similarity. Diamond OA shows a similar but less pronounced pattern, with modest advantages concentrated at high similarity levels and greater dispersion across the distribution.

The contrast is particularly sharp for body citations, where the separation between gold OA and the other models becomes clearly visible across a wide range of similarity values. Hybrid and bronze OA largely overlap with the closed-access reference throughout the distribution, suggesting that their prominence in patent citations does not translate into systematically stronger cognitive alignment. For publications cited in both the front section and the body, gold OA again displays the clearest distributional advantage, while diamond OA exhibits intermediate behavior and green OA remains close to the closed baseline.

Taken together, the ECDF analysis corroborates and generalizes the evidence from Figures 3 to 5. Differences across OA models are not driven by a small number of extreme observations nor by mean effects alone. Instead, fully open publication models, especially gold OA, exhibit a broad and systematic dominance in semantic proximity precisely in citation contexts associated with substantive use of scientific knowledge, providing strong distribution-level support for H3.

### 4.4. Multivariate analysis and robustness checks

To assess whether the distributional patterns documented above persist once key confounding factors are jointly controlled for, we estimate a multivariate linear regression model in which the dependent variable is the semantic similarity between patent and publication abstracts. The model includes indicators for OA status, disciplinary domain, and citation position (Front, Body, Both), as well as controls for publication year and field-weighted citation impact (FWCI). Closed access, Life Sciences, and Front citations serve as reference categories.

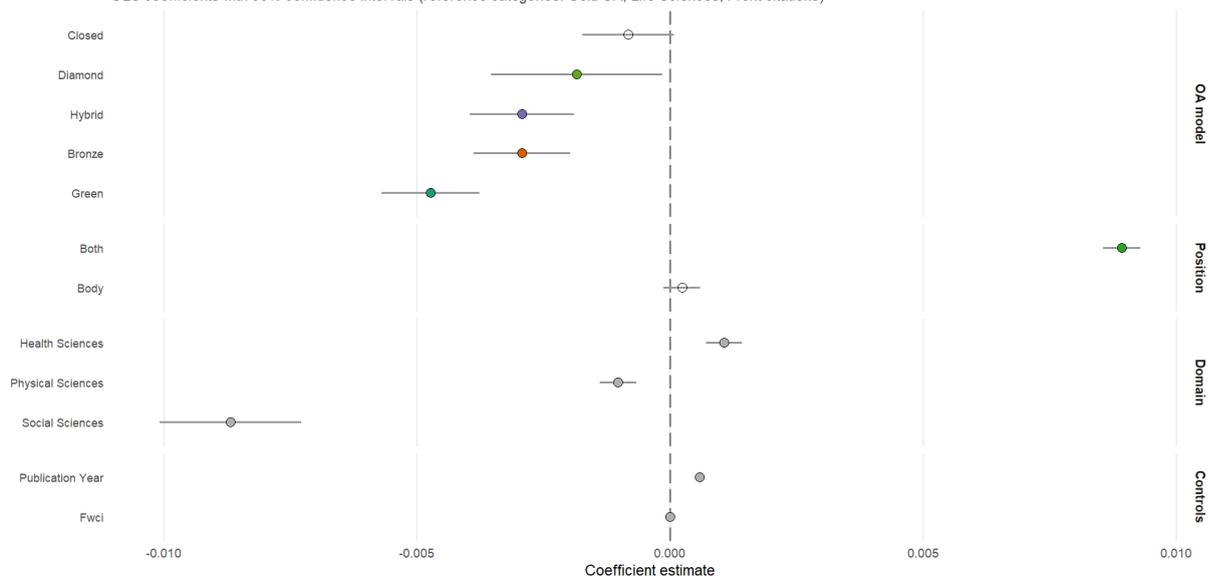

Figure 7. Multivariate determinants of semantic proximity between patents and publications
OLS coefficients with 95% confidence intervals (reference categories: Gold OA, Life Sciences, Front citations)

The results confirm that OA models differ significantly in their association with semantic proximity, even after accounting for domain, citation position, publication year, and scientific impact. Relative to gold OA (reference), green, bronze, and hybrid OA are associated with significantly lower semantic similarity, with effect sizes that are modest in absolute terms but highly statistically significant. Diamond OA also exhibits a negative coefficient relative to gold OA, though smaller in magnitude, indicating intermediate performance between fully open gold journals and more traditional OA models. The coefficient for closed access is not statistically different from gold OA at conventional levels, suggesting that the strongest contrast is not between open and closed per se, but between fully open models and partially or institutionally mediated OA models.

Citation position remains a powerful predictor of semantic proximity. Publications cited in both the front section and the body of patents display a substantially higher semantic similarity than those cited only in the front section, even after controlling for all other covariates. By contrast, body-only citations do not differ significantly from front citations once controls are included. This result is consistent with the interpretation that dual-position citations capture publications that play a more central and substantive role in the inventive process.

Disciplinary differences are also pronounced. Relative to Life Sciences, semantic proximity is significantly lower in Social Sciences and Physical Sciences, while Health Sciences exhibit higher similarity scores. These patterns align with well-documented differences in the codification of knowledge and the intensity of science–technology linkages across fields. As expected, publication year is positively associated with semantic similarity, reflecting the increasing textual convergence between scientific publications and patents over time, while FWCI is also positively and significantly related to semantic proximity, indicating that more influential scientific publications tend to be more closely aligned with patented technologies.

To sum up, the multivariate analysis corroborates the main conclusions drawn from the descriptive and distributional analyses. The dominance of gold OA in high-proximity patent–science links is not explained by disciplinary composition, citation position, recency, or scientific impact. Instead, it reflects a robust association between fully open publishing models and deeper cognitive alignment with technological invention, providing strong confirmatory evidence in support of H3 and reinforcing the interpretation that visibility and discoverability alone cannot account for the observed patterns.

## 5. Discussion

This paper examined how different OA publishing models shape the role of scientific knowledge in technological innovation, distinguishing between the selection of scientific publications into patent citations and their translation into technological content. Combining a large-scale non-patent reference data with semantic similarity measures and citation-position information, this analysis offers a more nuanced understanding of how scholarly communication infrastructures mediate science-technology linkages.

A central contribution of this study is to show that visibility-driven selection and cognitive alignment are partially decoupled. Consistent with a large literature using patent citations as indicators of science–technology linkages (Narin et al., 1997; Ahmadpoor et al., 2017), we find that patent citations disproportionately draw on publications disseminated through institutionally established and highly visible publishing channels. Hybrid and bronze OA publications, often associated with long-standing journals and publishers, are overrepresented among patent citations, even after controlling for disciplinary composition.

However, extending beyond citation counts, the semantic analysis reveals that this dominance does not translate into stronger cognitive alignment with patented technologies. On the contrary, publications published in fully OA journals (particularly gold OA, and to a lesser extent diamond OA) exhibit equal or higher semantic proximity to patents, especially when citations appear in the body of patent texts. These findings challenge interpretations that equate higher citation frequency with greater technological relevance and help reconcile long-standing debates about the meaning of patent citations (Meyer, 2000; Jaffe et al., 2000).

Therefore, the results suggest that patent citations reflect not only technological relevance but also the institutional organization of discoverability in scholarly communication.

The findings invite a reinterpretation of OA models not as simple access regimes, nor as proxies for journal quality, but as communication infrastructures embedded in broader information systems. Hybrid and bronze OA publications benefit from strong integration into indexing services, citation databases, and professional search routines used by inventors, patent examiners, and patent offices. These features enhance their discoverability and likelihood of citation, independently of the substantive content of the research.

Fully OA journals, by contrast, may disseminate research that is equally or more closely aligned with technological problem-solving, yet face structural disadvantages in terms of visibility and institutional recognition. This interpretation is consistent with recent work emphasizing the role of metadata quality, indexing, and platform integration in shaping knowledge diffusion beyond academia [Zhou et al., 2025]. Importantly, the results caution against reading underrepresentation in patent citations as evidence of lower relevance or quality.

The differentiation between citation positions within patents provides further insight into the mechanisms at play. Differences across OA models are weak or absent for front-section citations, which are often associated with formal disclosure requirements. By contrast, OA-related differences in cognitive alignment are strongest for citations embedded in the body of patent texts, where scientific knowledge is more likely to be substantively mobilized in the inventive process. This finding contributes to the literature that seeks to unpack the heterogeneous functions of patent citations (Criscuolo et al., 2008; Chen et al., 2017). It suggests that where a scientific publication is cited matters as much as whether it is cited, and that publishing models shape science–technology linkages primarily through selection mechanisms rather than through the depth of knowledge integration.

More broadly, the results help explain why empirical studies on OA and innovation have produced mixed findings. Analyses focusing on citation probabilities or counts capture selection effects driven by visibility and accessibility, whereas analyses concerned with technological relevance require measures that capture cognitive alignment. Without distinguishing these dimensions, the relationship between OA and innovation risks being mischaracterized.

## 6. Conclusion

This paper examined how different open OA publishing models shape the translation of scientific knowledge into technological innovation, moving beyond citation counts to analyze the substantive cognitive alignment between science and patents. Using large-scale data on non-patent references and multiple complementary approaches (structural comparisons, semantic similarity measures, distributional analyses, and multivariate models) we provide a comprehensive assessment of how publishing models condition the role of science in inventive activity.

Our results reveal a clear dissociation between visibility and substantive use. Patent citations disproportionately rely on publications disseminated through institutionally established and highly visible channels, particularly hybrid and bronze OA. However, this dominance in citation volume does not translate into stronger cognitive proximity to patented technologies. Across citation positions, distributional segments, and model specifications, publications published in fully open journals (especially gold OA, and to a lesser extent diamond OA) exhibit equal or higher semantic alignment with patents, with the strongest effects observed when citations signal substantive engagement with scientific knowledge.

These findings challenge the implicit assumption that access alone drives the technological relevance of science. Instead, they point to the importance of how publishing models are embedded in broader information infrastructures that shape discoverability, interpretability, and reuse. Fully open models appear to facilitate not only access, but also deeper integration of scientific content into technological development, while partially open models benefit primarily from inherited visibility within established publishing ecosystems.

From a policy perspective, the results suggest that maximizing the economic and technological returns to open science requires more than expanding OA mandates. Efforts to improve the discoverability, indexing, and integration of fully open journals within innovation-relevant information systems may be as important as lowering access barriers. Concretely, this implies strengthening standardized metadata requirements, promoting the systematic assignment of persistent identifiers, and ensuring that fully open publications are systematically harvested by major patent search platforms and professional prior-art databases. Public funders and research infrastructures could also prioritize investments in interoperable repositories and machine-readable full texts to reduce search frictions for both inventors and patent examiners. More broadly, the study underscores that publishing models are not neutral transmission channels: they structure which scientific knowledge is mobilized in innovation, and how deeply it is absorbed.

## 7. Limitations

This study relies on the OA classification provided by OpenAlex, which distinguishes between gold, hybrid, bronze, green, and closed access based on publisher licensing and repository availability (see OpenAlex documentation). While this classification is widely used and appropriate for large-scale analyses, it reflects the availability of different versions of a publication rather than the specific version accessed by users.

In particular, publications classified as green OA may also exist in closed-access versions at the publisher. As a result, the data do not allow us to determine whether inventors or patent examiners relied on the repository version or the publisher version. This limitation is inherent to large-scale bibliometric datasets and should be interpreted as a source of measurement imprecision rather than a systematic bias.

Importantly, this ambiguity is unlikely to affect the main comparative patterns identified in this study, as it primarily concerns a subset of publications available under multiple access routes (e.g., green and closed, or green and hybrid). The results should therefore be interpreted as reflecting differences across OA categories as defined in widely used data infrastructures, rather than exact modes of access in individual cases.

More broadly, OA categories are best understood as features of publication records within scholarly communication systems, which shape visibility and dissemination conditions, rather than direct observations of how knowledge is accessed in practice.

## 8. Appendices

**A1. Descriptive Statistics of Data Distribution and Semantic Similarity**

Based on 2,015,841 "patent-paper" citation pairs and utilizing text vectors generated by the SPECTER 2 model, this study conducted a matching analysis on 632,756 scientific paper abstracts and 413,586 patent abstracts. To investigate the impact of OA types on the technical-scientific semantic linkage, we performed a detailed statistical analysis of the dataset's composition and the text similarity across groups (Table 1).

(1) Sample Composition and Distribution Imbalance: regarding the distribution of OA types among citation pairs, the sample exhibits significant imbalance. Non-OA (closed) papers are overrepresented, totaling 1,067,232 pairs (accounting for approximately 52.9%). Among OA categories, bronze (388,650 pairs) and green (357,500 pairs) hold larger shares, followed by hybrid. In contrast, sample sizes for gold (53,454 pairs) and diamond (2,531 pairs) are relatively small. This quantitative disparity objectively reflects that the existing stock of literature cited by patents is still dominated by traditional subscription and archival models.

(2) Central Tendency of Semantic Similarity: Table 1 presents the statistics of semantic similarity between patents and papers under different OA types. Overall, the mean similarity (Mean) for all groups falls within the high range of 0.88 to 0.89, with medians generally slightly higher than means. This indicates that patents and their cited papers maintain an extremely high semantic relevance overall, validating the substantive effectiveness of the citation relationship. Regarding specific group differences, gold OA demonstrates the highest semantic consistency, with a mean similarity of 0.8921; diamond OA follows closely at 0.8905. In comparison, the mean similarity for band green OA is slightly lower (approx. 0.886). This suggests that papers published via gold or diamond routes—which require peer review—exhibit a closer "semantic distance" to the technical content of the patents citing them.

(3) Analysis of Statistical Robustness and Dispersion Characteristics: It is worth noting that although there are huge differences in sample sizes between different OA types (e.g., the closed group is about 400 times the size of the diamond group), this does not affect the statistical validity of the comparison of means between groups. First, the variance across groups shows high consistency, with values stable around 0.0007. This indicates that the degree of data dispersion has not been distorted by the disparity in sample sizes, and the distribution shape of the data across groups is similar and stable. Second, even the diamond group with the smallest sample size (N=2,531) far exceeds the threshold for a large sample in a statistical sense, sufficient to guarantee the convergence of the mean estimate based on the Law of Large Numbers. Furthermore, extreme value analysis shows that the minimum similarity value of the diamond group (Min = 0.7637) is significantly higher than that of the closed group (Min = 0.7068). Combined with the aforementioned variance consistency, this implies that high-quality OA models may filter out low-relevance citations at the lower bound, rather than the differences being attributed to random errors caused by sample size fluctuations.

Table 1. Descriptive Statistics of Data Distribution and Semantic Similarity

| OA Type | Paper Count | Patent-Paper Pairs | Max Similarity | Min Similarity | Median Similarity | Mean Similarity | Variance |
|---|---|---|---|---|---|---|---|
| Bronze | 116,028 | 388,650 | 0.9890 | 0.7194 | 0.8874 | 0.8857 | 0.0007 |
| Closed | 363,614 | 1,067,232 | 0.9940 | 0.7069 | 0.8893 | 0.8875 | 0.0007 |
| Diamond | 1,245 | 2,531 | 0.9536 | 0.7637 | 0.8934 | 0.8905 | 0.0007 |
| Gold | 22,513 | 53,454 | 0.9861 | 0.7281 | 0.8938 | 0.8921 | 0.0007 |
| Green | 87,653 | 357,500 | 0.9902 | 0.7431 | 0.8873 | 0.8859 | 0.0007 |
| Hybrid | 41,703 | 146,474 | 0.9764 | 0.7249 | 0.8887 | 0.8867 | 0.0007 |

## A2. Robustness checks

This appendix includes supplementary analyses exploring the sensitivity of the results to alternative specifications and sample restrictions. In particular, we report additional regression models and descriptive statistics that support the robustness of the main findings. These analyses reinforce the conclusion that the associations between OA models and patent–science proximity are not driven by disciplinary composition or citation position alone.

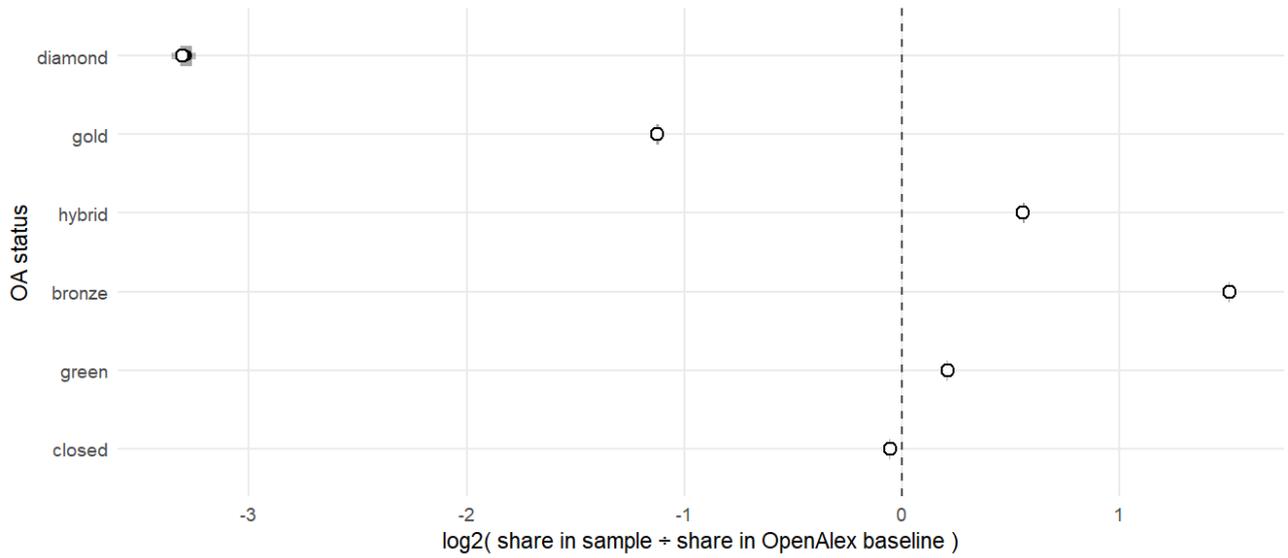

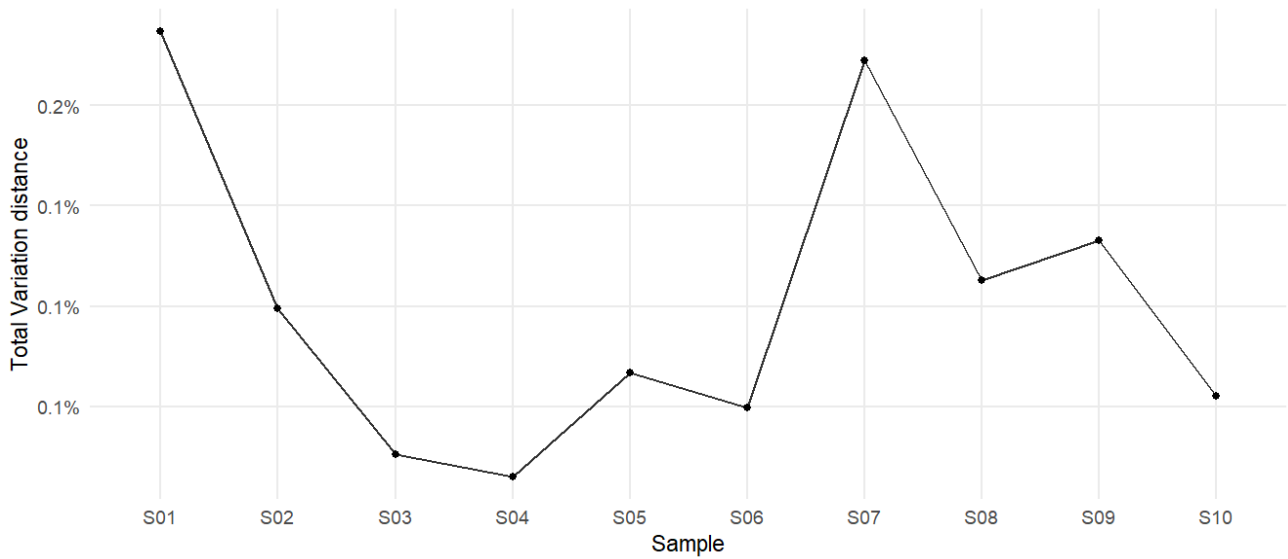

## A3. Semantic scores distribution by OA status

**Figure A3** displays kernel density (ridgeline) plots of semantic similarity scores between patents and cited scientific publications by OAs model and citation position.

These plots provide an intuitive visualization of the full distributions underlying the average and quantile-based results reported in the main text.

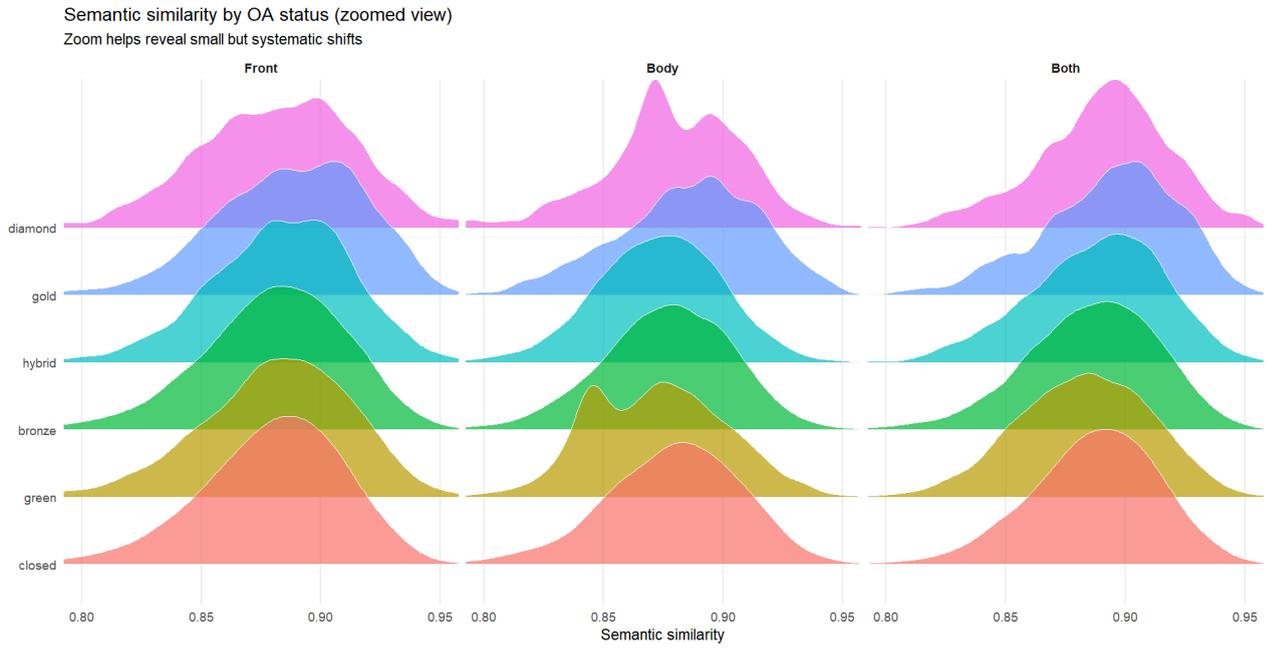

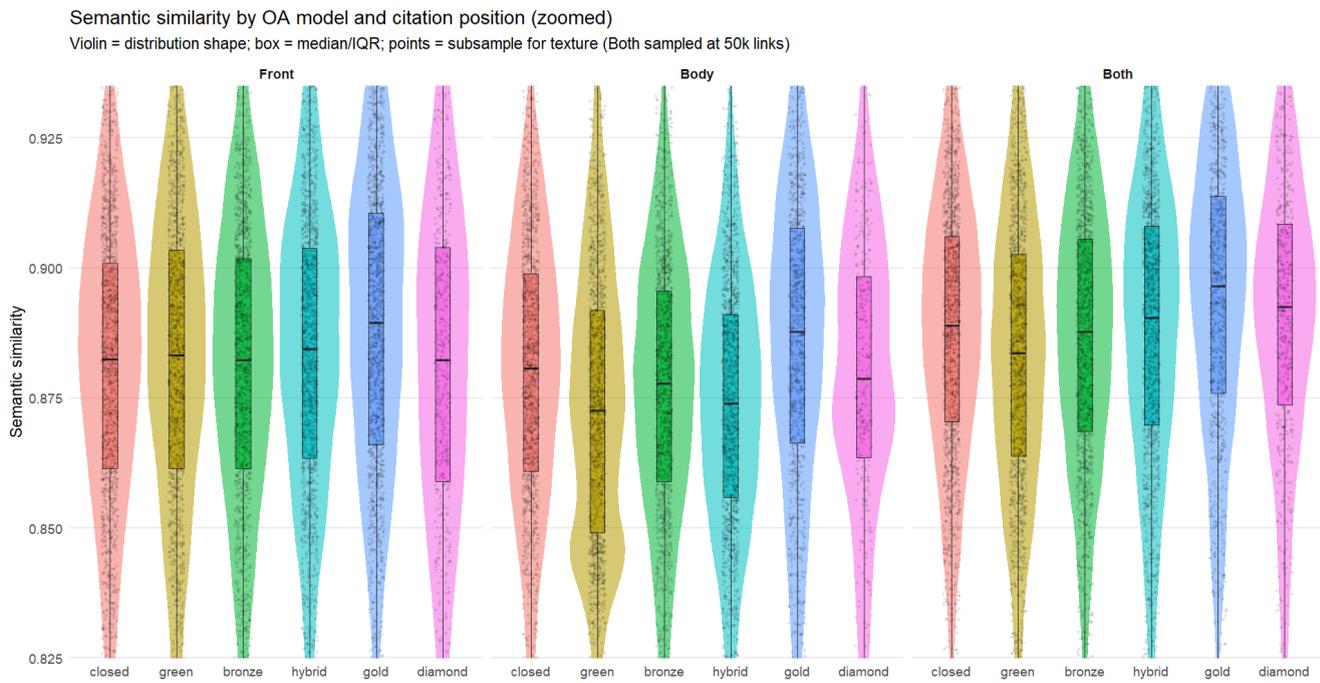

## A4. Regressions with interactions

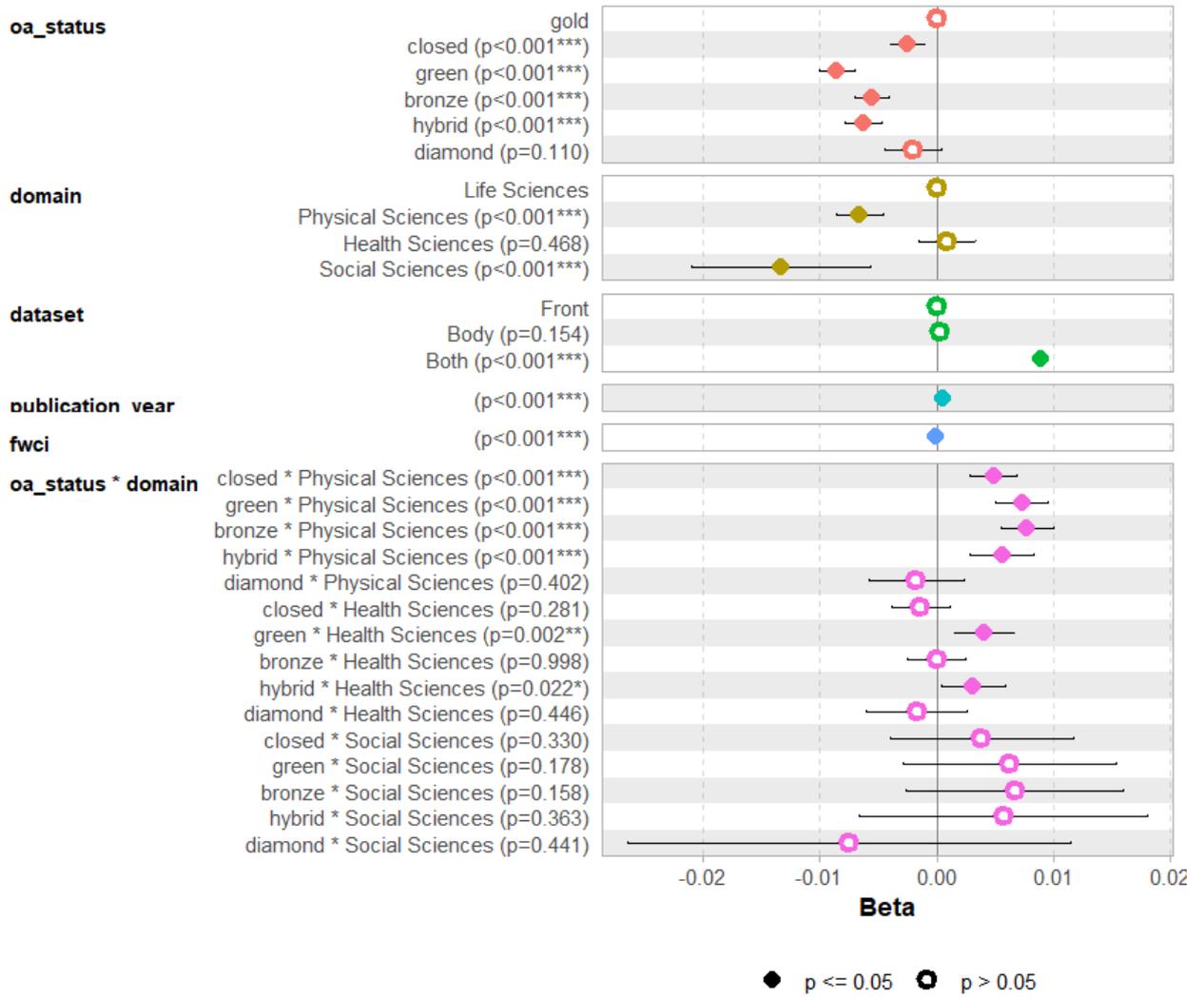

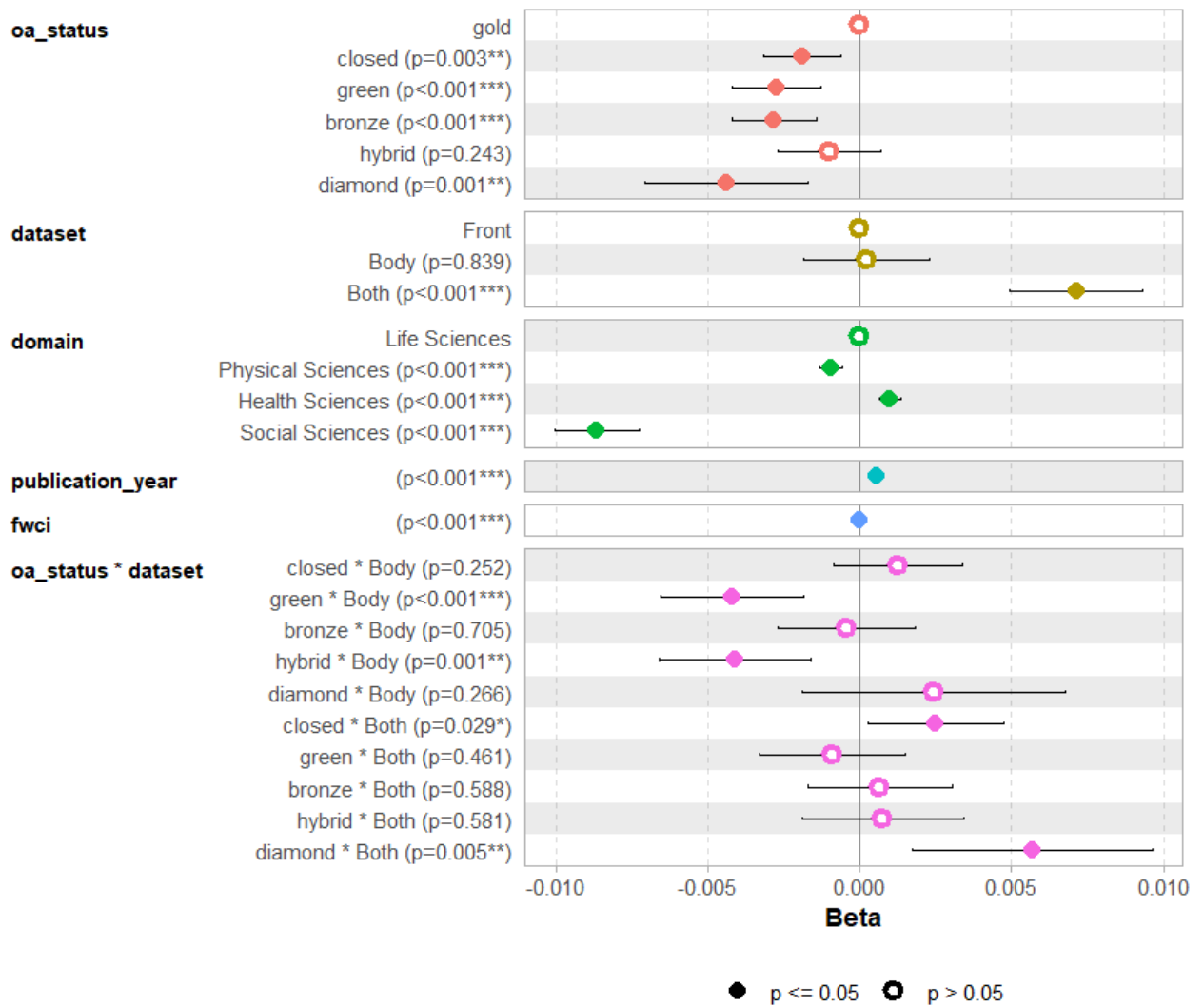